\documentclass[conference,10pt]{IEEEtran}
\IEEEoverridecommandlockouts
\usepackage{graphicx}
\usepackage{amsmath}
\usepackage{amssymb}
\usepackage{amsfonts}
\usepackage{amsopn,theorem}
\newtheorem{theorem}{\bf Theorem}

\newtheorem{remark}{\bf Remark}

\def\tR{\tilde{R}}

\def\n{\nonumber\\}
\def\be{\begin{equation}}
\def\ee{\end{equation}}
\def\bes{\begin{equation*}}
\def\ees{\end{equation*}}
\def\beq{\begin{eqnarray}}
\def\eeq{\end{eqnarray}}
\def\beqs{\begin{eqnarray*}}
\def\eeqs{\end{eqnarray*}}

\def\ma{{\mathcal A}}

\def\ms{{\mathcal S}}

\def\mx{{\mathcal X}}

\def\tv#1{\left\|#1\right\|_1}
\def\apx#1{\stackrel{#1}{\approx}}

 \def\clap#1{\hbox to 0pt{\hss#1\hss}}

\usepackage{tikz}

\usetikzlibrary{arrows,shapes.symbols}
\usetikzlibrary{through}
\usetikzlibrary{fit}					
\usetikzlibrary{backgrounds}	

\begin{document}

\author{Amin~Gohari, Mohammad~Hossein~Yassaee and Mohammad~Reza~Aref\\Information Systems and Security Lab (ISSL),\\Sharif University of Technology, Tehran, Iran,\\
E-mail: aminzadeh@sharif.edu, yassaee@ee.sharif.edu, aref@sharif.edu.
\thanks{\scriptsize \noindent This work was partially supported by Iranian National Science Foundation (INSF) - cryptography chair.}
}
\title{Secure Channel Simulation}

\maketitle
\begin{abstract}
In this paper the Output Statistics of Random Binning (OSRB) framework is used to prove a new inner bound for the problem of secure channel simulation. Our results subsume some recent results on the secure function computation. We also provide an achievability result for the problem of simultaneously simulating a channel and creating a shared secret key. A special case of this result generalizes the lower bound of Gohari and Anantharam on the source model to include constraints on the rates of the public discussion.
\end{abstract}
\section{introduction}
Output statistics of random binning \cite{OSRB} is a new framework for proving achievability results. In this paper we use this framework to extend the secure function computation of \cite{Tyagi} for the case of two users, where two users are observing i.i.d. repetitions of $X_1$ and $X_2$ and would like to construct i.i.d. repetitions of $Y=f(X_1, X_2)$ after interactively exchanging messages on a public channel. I.i.d. repetitions of the function $Y$ has to remain nearly independent of the messages exchanged. It was shown in \cite{Tyagi} that this is possible if and only if $H(Y)<I(X_1;X_2)$. This work was further generalized in \cite{Tyagi2}.
We extend the achievability part of the existing results by assuming that there is an eavesdropper who has access to i.i.d. repetitions of $Z$. Further in our model the two party want to generate i.i.d. repetitions of $Y_1$ and $Y_2$ where $Y_1$ and $Y_2$ are not necessarily functions of $X_1$ and $X_2$; they are jointly distributed with $X_1$, $X_2$ and $Z$ according to some arbitrary $p(y_1,y_2|x_1,x_2)p(x_1,x_2,z)$. We demand a reliable generation of $Y_1^n$ and $Y_2^n$ meaning that the total variation distance between the pmf of the generated $(Y_1^n, Y_2^n, X_1^n, X_2^n, Z^n)$ and the i.i.d. pmf must go to zero asymptotically as $n$ goes to infinity. Further, the public discussion must reveal no new information to Eve about an $S^n$, created by passing $(Y_1^n, Y_2^n, X_1^n, X_2^n, Z^n)$ of the code through $n$ copies of the channel $p(s|x_1, x_2, y_1, y_2, z)$. A special case of interest is when $S=(Y_1, Y_2)$ meaning that we would like to keep the generated rv's hidden from Eve. In our model we further assume rate limited public discussion and a preshared secret key at rate $R_0$. Lastly we provide an achievability result for the problem of simultaneously simulating a channel and creating a shared secret key. A special case of this result generalizes the lower bound of Gohari and Anantharam on the source model \cite{SKpaper} to include constraints on the rates of the public discussion.

%
The paper is organized as follows: in Section \ref{section:review} we review the output statistics of random binning technique at some length. In Section \ref{section:ChannelSimuation} we discuss our new inner bound for the secure channel simulation problem. In Section \ref{section:KeyGeneration} we discuss simultaneous simulation of a channel and generation of a secret key.

\par \textbf{Notation}: All random variables are taking values in finite sets. We use $[1:r]$ to denote the set $\{1,2,3,\dots, r\}$, $X_{\ms}$ to denote $(X_j:j\in\ms)$ and $p^U_{\ma}$ to denote the uniform distribution over the set $\ma$. Given a natural number $i$, $(i)_2$ is 1 if $i$ is odd, and is 0 if $i$ is even. 
The total variation between two pmf's $p$ and $q$ on the same alphabet $\mx$, is defined by $\tv{p(x)-q(x)}:=\frac{1}{2}\sum_x|p(x)-q(x)|$.

\section{Review of Output Statistics of Random Binning}\label{section:review}
To illustrate the main ideas behind the OSRB technique, we begin by two examples, each of which connects a source coding problem to a channel coding problem. Our discussion is at an intuitive level; see \cite{OSRB} for a rigorous treatment.

The first example connects Wyner's wiretap channel \cite{Wyner} to the one-way source model key agreement problem \cite{Ahlswede}. Consider the source model key agreement problem: Alice, Bob and Eve have access to i.i.d. repetitions of $X^n, Y^n$ and $Z^n$ respectively, distributed according to $\prod_{i=1}^np(x_i,y_i,z_i)$. It is known that the key rate $I(X;Y)-I(X;Z)$ is achievable (when $I(X;Y)-I(X;Z)>0$). To obtain this rate, Alice sends the Slepian-Wolf (SW) index of $X^n$ to Bob (at rate $H(X|Y)+\epsilon$) over a public channel. Then Alice constructs the key $M$ by binning $X^n$ into $2^{n{(I(X;Y)-I(X;Z)-\epsilon)}}$ bins (this binning is independent of the SW binning). If we denote the public message by $B$ and the key by $M$, the following hold: both $B$ and $M$ are random bin indices of $X^n$, and the key $M$ is nearly independent of $(B,Z^n)$. Thus there is an instance of $B=b$ such that conditioned on $B=b$ the following two properties hold: $M$ is nearly independent of $Z^n$, and Bob can recover the key $M$ with high probability (conditioned on $B=b$). Since $B$ is a function of $X^n$, we have the factorization $p(x^n,y^n,z^n|b)=p(x^n|b)p(y^n,z^n|x^n)$. In other words conditioning on $B=b$ only changes the marginal distribution of $X^n$ but leaves the channel from $X^n$ to $(Y^n,Z^n)$, i.e. $p(y^n,z^n|x^n)$, undisturbed. Further $p(m,z^n|b)\simeq p(m)p(z^n|b)$ and Bob can almost recover $M$ from $Y^n$ conditioned on $B=b$. The joint distribution of these random variables (conditioned on a fixed $B=b$) can be used to construct a code for secure transmission over a wiretap channel $p(y,z|x)$. We interpret $M$ as the message to be transmitted. Since $M$ is nearly independent of $B$, conditioning on $B=b$ does not change its marginal distribution (thus it is still uniform over a set of size $2^{n{(I(X;Y)-I(X;Z)-\epsilon)}}$). Further conditioned on $B=b$, the message $M$ is nearly independent of $Z^n$ and can be recovered from $Y^n$. Lastly $p(y^n,z^n|x^n,b)=p(y^n,z^n|x^n)$. This shows that the rate $I(X;Y)-I(X;Z)$ is achievable for the wiretap problem. It is not difficult to modify this proof to show that $\max_{p(u,x)}I(U;Y)-I(U;Z)$ is also achievable for the wiretap channel problem (and indeed this is the capacity region).

Next, consider the problem of sending a message $M$ of rate $R$ over the channel $p(y|x)$. The input distribution $p(x^n)$ is uniform over the $2^{nR}$ codewords, thus it is \emph{not} i.i.d. .  However Shannon's idea of generating a random codebook makes the input distribution i.i.d. . Shannon noted that granting a preshared randomness between the encoder and decoder (denoted by $B$ and independent of the message $M$) does not increase the capacity of the channel (see the top diagram of Fig. 1). However the encoder and decoder can use this common randomness to generate an i.i.d. random codebook. Once the random codebook is generated at both the encoder and the decoder, a codeword is chosen according to the value of $M$ and is transmitted over the channel. Thus we have an encoder $X^n(M,B)$ and a decoder $\hat{M}(Y^n,B)$. Since the probability of error is the average of that over all realizations of $B$, one can find $b$ such that $X^n(M,B=b)$ and $\hat{M}(Y^n, B=b)$ form appropriate encoder and decoder. The input $X^n(M,B)$ is i.i.d., although $X^n(M,B=b)$ is not so. Now, note that the joint pmf $P_{M, B ,X^n}=P_MP_BP_{X^n|M,B}$ can also be written as $P_{X^n}P_{M,B|X^n}$. This is as if we generate an i.i.d. $X^n$ and pass it through a virtual reverse encoder $P_{M,B|X^n}$ to generate $M$ and $B$. This is depicted in the bottom diagram of Fig. 1 where we have changed the direction of arrows to reflect this change of order. In this interpretation we are starting from an i.i.d. $X^n$ and $Y^n$ according to $\prod_{i=1}^np(x_i,y_i)$. Random variable $B$ is now a (public) message transmitted from the transmitter to the receiver. We can view it as the Slepian-Wolf message from $X^n$ to $Y^n$. Once the decoder has recovered $X^n$ it can recover $M$, if $M$ is a function of $X^n$. Now we are ready to create the source coding counterpart. We take some arbitrary $p(x)$ and generate $n$ i.i.d.\ copies of $X^n$ and $Y^n$ according to $p(x)p(y|x)$. We then construct $B$ and $M$ as random partitions (binnings) of $X^n$. Random variable $B$ is a SW index of size $n(H(X|Y)+\epsilon)$. It enables the receiver to recover $X^n$ with high probability. Thus, the receiver can recover $M$. Next we see that in the channel coding side, $M$ and $B$ are independent and $M$ is uniform. Thus we are looking for constraints that make bin indices $B$ and $M$ of an i.i.d. $X^n$ independent, and $M$ uniform. It turns out that as long as $\log|\mathcal{B}|+\log|\mathcal{M}|<nH(X)$, rv's $B$ and $M$ are independent, and $M$ is uniform. This holds for instance if $|\mathcal{M}|<2^{I(X;Y)-2\epsilon}$, giving us the rate $I(X;Y)-2\epsilon$. To go back to the channel coding problem we look at the $P_{X^n,M,B}$ imposed by $M$, $B$ and $X^n$. Next we take $P_{X^n|M,B}$ and use it in the channel coding setup of Fig. 1. To get away with shared randomness $B$, we observe that we still have the property that $p(y^n|x^n,B=b)=p(y^n|x^n)$ and $p(m|B=b)\apx{}p(m)$ meaning that $X^n(M,B=b)$ and $\hat{M}(Y^n, B=b)$ are legitimate choices as the encoder and decoder; we are done.

Observe the secrecy flavor of the source coding side of the problem: we start from i.i.d. repetitions of $X^n, Y^n$; we can interpret $B$ as a public message, and $M$ as a secret key which is independent of $B$. This is an instance of the source model SK generation problem.

\begin{figure}
\begin{center}
\begin{tikzpicture}[scale=1.1,>=stealth']
\def\bscale{.5}
\tikzstyle{enc}=[scale=.8,draw=black, text width=2.4em, rounded corners,
    text centered, minimum height=2em,draw=blue,
    ]
    \tikzstyle{ch}=[scale=.8,draw=black, minimum width=2.6em, rounded corners,
    text centered, minimum height=6em,draw=blue,
    ]
\tikzstyle{ann} = [above, text width=2em]
\tikzstyle{dec} = [enc, text width=2.4em,
    minimum height=2em,fill=gray!70!white,draw=black]
\node (ch) [enc] {{$p_{Y|X}$}};
\path (ch.180)+(-1.5,0) node[] (xn) {\small \textcolor{black}{$X^n(M,\color{red}B\color{black})$}};
\path (ch.180)+(-3,0) node[enc] (enc) {$\mathsf{Enc}$};
\path (enc.180)+(-.5,0) node (tx){\small $M$};
\path (enc.90)+(0,.5) node (txx){\small \color{red}{$B$}\color{black}};
\path (ch.0)+(1.5,0) node [enc](dec){$\mathsf{Dec}$};
\path (dec)+(1,0) node (d1){\small $\widehat{M}$};
\draw [->] (tx)--(enc);
\draw [->,red] (txx)--(enc);
\draw[->,red] (txx)-|(dec);
\path [draw,->] (enc)--(xn);
\path [draw,->] (xn)--(ch);
\path [draw,->] (ch)--node[above]{\small \textcolor{black}{$Y^n$}}(dec);
  \coordinate  (d) at (tx.180|-txx.90);
\end{tikzpicture}\label{fig:reverse-encoder}
\end{center}
\vspace{0.2cm}
%
%
\begin{center}
\begin{tikzpicture}[scale=1.32,>=stealth']
\def\bscale{.5}
\tikzstyle{enc}=[scale=.8,draw=black, text width=2.4em, rounded corners,
    text centered, minimum height=2em,draw=blue,
    ]
    \tikzstyle{ch}=[scale=.8,draw=black, text width=3.2em, rounded corners,
    text centered, minimum height=2em,draw=blue,
    ]
\tikzstyle{ann} = [above, text width=2em]
\tikzstyle{dec} = [enc, text width=2.4em,
    minimum height=2em,fill=gray!70!white,draw=black]
\node (ch) [enc] {{$p_{Y|X}$}};
\path (ch.180)+(-.8,0) node[] (xn) {\small \textcolor{black}{$X^n$}};

{\path (ch.180)+(-1.8,0) node[ch] (enc) {$P_{MB|X^n}$};}
\path (enc.180)+(-.5,0) node (tx){\small $M$};
\path (enc.90)+(0,.5) node (txx){\small \color{red}$B$\color{black}};
\path (ch.0)+(1.5,0) node [enc](dec){$\mathsf{Dec}$};
\path (dec)+(1,0) node (d1){\small $\widehat{M}$};
{\draw [<-,thick] (tx)--(enc);
\draw [<-,red,thick] (txx)--(enc);
\path [draw,<-,thick] (enc)--(xn);
}
\draw[->,red] (txx)-|(dec);
\path [draw,->] (xn)--(ch);
\path (ch.0)+(.5,0) node (yn){\small \textcolor{black}{$Y^n$}};
\path [draw,->] (ch)--(yn)--(dec);
  \coordinate  (d) at (tx.180|-txx.90);

   \begin{pgfonlayer}{background}
\draw[very thick, dashed,green!50!black,fill=yellow!20!white] (xn.0)+(0,-.3) rectangle (d);
    \end{pgfonlayer}


\end{tikzpicture}
\caption{(Top) Point-to-point channel with preshared randomness $B$ to generate a random codebook. (Bottom) The corresponding source coding problem: reversing the order of generating rv's in the box.
}
\end{center}
\vspace{-0.8cm}
\end{figure}
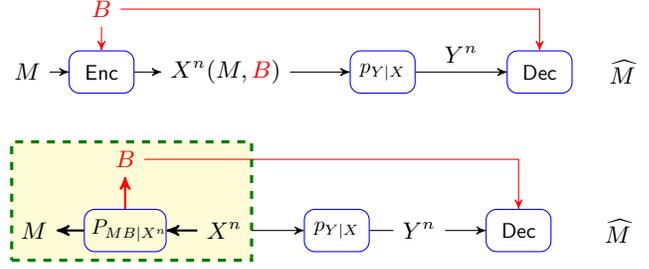

The OSRB framework is a systematic way of converting channel coding problems into source coding problems (the above examples show how that can happen). The advantage of the conversion is that in the source coding side of the problem we only have \emph{one} copy of the random variables, e.g. in the point to point example we start from a \emph{single} i.i.d. copy of $X^n$, $Y^n$; all the other rv's (i.e. $M$ and $B$) are random bins of these i.i.d. rvs. However if we were to directly attack the channel coding problem, we had to create a codebook of size $2^{nR}$ containing \emph{lots of} $x^n$ sequences. This conversion is useful in problems involving multi-round interactive communication with several auxiliary random variables (e.g. the problem studied in this paper) where it is desirable to have just a \emph{single} i.i.d. repetition of all the original and auxiliary random variable (rather than having many i.i.d. copies of these random variables related to each other through superposition or Marton coding type structures). Once we take a single i.i.d. copy, all the messages and preshared randomness (such as $B$) can be constructed as random bins of these i.i.d. rv's. Traditional coding techniques start with the messages and then create the many codewords. Here we are reversing the order by starting from a single i.i.d. copy of the original and auxiliary rv's, and constructing the messages as bin indices afterwards. And this can simplify representing the codebook construction and analyzing its probability of success. For instance while the traditional framework considers superposition coding and Marton coding as distinct coding constructions, in the new framework the two constructions are nothing but two different ways of specifying the set of i.i.d. rv's we are binning. Thus the new framework unifies the two coding strategies, for it \emph{only} uses random binning.

In the traditional framework we need to count the size of typical sets; this is generally done via covering and packing lemmas. However in the OSRB framework we need to find two sets of conditions: one set of conditions for Slepian-Wolf decoders to succeed and another set implying independence of certain random bin indices. Thm. 1 of \cite{OSRB} provides sufficient conditions for the latter. This change from counting typical sequences to working with output statistics of random binnings provides a framework to prove results under a strong notion of security conveniently. This is partly due to the fact that OSRB brings the randomness of random codebook generation from the background into the foreground as an explicit rv (e.g. $B$ in the above example), or a set of rv' s.

\section{Secure Channel Simulation By Two Terminals} \label{section:ChannelSimuation}
We begin with the formulation of the problem without any secrecy constraints as in \cite{me2}:

\subsection{Channel Simulation with no secrecy constraints}
Assume that Alice and Bob observe i.i.d.\ repetitions of two random variables $X_1$ and $X_2$ respectively, and would like to generate i.i.d.\ repetitions of rv's $Y_1$ and $Y_2$ respectively. Random variables $X_1, X_2, Y_1, Y_2$ are jointly distributed according to a given $p(x_{[1:2]})p(y_{[1:2]}|x_{[1:2]})$. Alice and Bob are also provided with shared randomness at a rate $R_0$. The two parties can interactively talk to each other over $r$ rounds as they wish; the only constraints are that the total communication rate from Alice to Bob is bounded from above by $R_{12}$ and the total communication rate from Bob to Alice is bounded from above by $R_{21}$. The question is for which values of $(R_0, R_{12}, R_{21})$ the pmf $p(x_{[1:2]},y_{[1:2]})$ can be asymptotically achieved; i.e. for every $\epsilon>0$ there is a sequence of $(n,\epsilon)$ codes that results in $\tilde{p}(x^n_{[1:2]},y^n_{[1:2]})$ satisfying the following for large $n$
\begin{align}\label{eq:e-c}
\tv{\tilde{p}(x^n_{[1:2]},y^n_{[1:2]}) - \prod_{i=1}^n p(x_{[1:2],i},y_{[1:2],i})}\leq \epsilon.
\end{align}
\begin{remark} When $Y_1$ and $Y_2$ are deterministic functions of $X_1$ and $X_2$, the problem would be that of finding two functions via interactive communication.\end{remark}

\begin{theorem} [Theorem 1 of \cite{me2}]\label{thm:thmme2}
The simulation rate region is the set $\ms(r)$ of all non-negative rate tuples $(R_0,R_{12},R_{21})$, for which there exists $p(f_{[1:r]}|x_{[1:2]},y_{[1:2]})\in T(r)$ such that
\begin{align}
R_{12}&\ge I(X_1;F_{[1:r]}|X_2),\label{eqT1}\\
R_{21}&\ge I(X_2;F_{[1:r]}|X_1),\label{eqT2}\\
R_0+R_{12}&\ge I(X_1;F_{[1:r]}|X_2)+I(F_1;Y_{[1:2]}|X_{[1:2]}),\label{eqT3}\\
R_0+R_{12}+R_{21}&\ge I(X_1;F_{[1:r]}|X_2)+I(X_2;F_{[1:r]}|X_1)\n&\qquad +I(F_{[1:r]};Y_{[1:2]}|X_{[1:2]}),\label{eqT4}
\end{align}\normalsize
where $T(r)$ is the set of $p(f_{[1:r]}|x_{[1:2]},y_{[1:2]})$ satisfying
\begin{align}
F_i-&F_{[1:i-1]}X_1-X_2, \ \mbox{if $i$ is odd,} \n
F_i-&F_{[1:i-1]}X_2-X_1, \ \mbox{if $i$ is even,} \n
Y_1-&F_{[1:r]} X_1-X_2Y_2,~~Y_2-F_{[1:r]} X_2-X_1Y_1.
\end{align}\normalsize
\end{theorem}

\begin{remark} The non-symmetric equation \eqref{eqT3} is due to the fact that the region is for a finite $r$ rounds of communication, with the first party starting the communication. The region would have been symmetric if the region was for infinite rounds of communication (i.e. $r\rightarrow \infty$).\end{remark}

To prove this theorem in \cite{me2}, we take some arbitrary $p(f_{[1:r]}|x_{[1:2]}, y_{[1:2]})\in T(r)$. We start from the source coding side of the problem where only a single i.i.d. copy of $(F_1^n,\cdots, F_r^n, X_{[1:2]}^n, Y_{[1:2]}^n)$ is created. The messages to be communicated in each stage $K_i$, the preshared randomness variables $B_i$, and the actual real shared randomness $\omega$ (of rate $R_0$) are created as bin indices of these i.i.d. variables in the following way: $B_1$, $K_1$ and $\omega$ are bin indices of three independent binning of $F_1^n$. Rv's $B_i$ and $K_i$ are bin indices of two independent binnings of $(F_1^n,\cdots,F_i^n)$. The alphabet sizes of $\omega$, $K_i$ and $B_i$ are $2^{nR_0}$, $2^{nR_i}$ and $2^{n\tR_i}$ respectively. Just as in the point to point case, there are going to be some constraints for the Slepian-Wolf decodings to work (similar to the point to point condition of $Y^n$ and $B$ being sufficient to recover $X^n$), and some constraints for independence of the bin indices (similar to the point to point condition of $B$ and $M$ being nearly independent) to allow us reverse the encoders and go from the source coding side to our original problem. We report the list of these conditions from \cite{me2}.
\begin{enumerate}
\item \emph{Reliability of SW decoders:}
\begin{align*}
R_1+R_0+\tR_1&\ge H(F_1|X_2),\\
\ R_i+\tR_i&\ge H(F_i|X_{(i+1)_2}F_{[1:i-1]})~~~\forall i\in[2:r].
\end{align*}\normalsize
where $(i)_2$ was defined at the end of introduction.
\item \emph{Independence constraints:}
\begin{align*}
R_0+\tR_1&< H(F_1|X_1),\\
\tR_i&<H(F_i|X_{(i)_2}F_{[1:i-1]})~~~\forall i\in[2:r],\\
\sum_{t=1}^i \tR_t&<H(F_{[1:i]}|X_{[1:2]}Y_{[1:2]})~~~\forall i\in[1:r].
\end{align*}\normalsize
\end{enumerate}
A Fourier-Motzkin elimination on the above constraints gives the region given in Thm. \ref{thm:thmme2}.
To intuitively understand the reliability of SW decoders constraints, note that common randomness $\omega$, $B_1$ and $K_1$ are random bin indices of $F_1^n$ created by Alice. Bob needs a rate of $H(F_1|X_2)$ from Alice to decode $F_1^n$ (and use it to create $F_2^n$ for the next round). This corresponds to the first SW constraint. Other SW constraints are similar with $B_i$ and $K_i$ serving as the random bin indices of $F_i^n$.

The first two independence constraints ensure that $B_{[1:r]}$, $\omega$ and $X_{[1:2]}^n$ are mutually independent: the first condition implies that $B_1$, $\omega$ and $X_{[1:2]}^n$ are mutually independent, and the second constraint implies that $B_i$ is nearly independent of $(B_{[1:i-1]}, \omega, X_{[1:2]}^n)$. To see this observe that the first independence constraint correspond to $B_1$ and $\omega$ being nearly mutually independent of each other and of $X_1^n$ (thus also independent of $X_{[1:2]}^n$ because of the Markov chain $F_1-X_1-X_2$ and the fact that $B_1$ and $\omega$ are bins of $F_1^n$). The second independence constraint implies that $B_i$ is nearly independent of $X_{(i)_2}^nF_{[1:i-1]}^n$. Because $B_i$ is a bin index of $F_i^n$ and because of the Markov chain $F_i-X_{(i)_2}F_{[1:i-1]}-X_{(i+1)_2}$, $B_i$ will be nearly independent of $X_{[1:2]}^nF_{[1:i-1]}^n$. Next since $B_{[1:i-1]}$ and $\omega$ are functions of $F_{[1:i-1]}^n$, $B_i$ will be nearly independent of $(B_{[1:i-1]}, \omega, X_{[1:2]}^n)$. Finally, the last independence constraint implies that $B_{[1:r]}$ is nearly mutually independent of $X^n_{[1:2]}Y^n_{[1:2]}$. Thus conditioning on a certain instance of $B_{[1:r]}=b_{[1:r]}$ does not disturb the joint pmf of $X^n_{[1:2]}Y^n_{[1:2]}$.

\subsection{Channel Simulation with an eavesdropper} We consider an eavesdropper (Eve) who is observing i.i.d. copies of $Z$, jointly distributed with $X_1$, $X_2$. We assume that Alice and Bob want to generate i.i.d.\ repetitions of $Y_1$ and $Y_2$ (within a vanishing total variation distance) jointly distributed with $X_1, X_2, Z$ according to a given $p(x_1, x_2, z)p(y_1, y_2\vert x_1, x_2)$. Meanwhile they want to make sure that the public discussion reveals no new information to Eve about an $S^n$, created by passing $(Y_1^n, Y_2^n, X_1^n, X_2^n, Z^n)$ of the code through $n$ copies of the channel $p(s|x_1, x_2, y_1, y_2, z)$. We assume that Alice and Bob are provided with a preshared secret key of rate $R_0$.

Public communications are rate constrained by $R_{12}$ and $R_{21}$ as before. The secrecy constraint is
$$\lim_{n\rightarrow\infty}\left|I(S^n;Z^n,K_1,\cdots,K_r)-nI(S;Z)\right|=0$$
over a sequence of codes where $K_1$, $K_2$, ..., $K_r$ are the messages exchanged during the $r$ rounds of interactive communication. Observe that we are using a strong notion of secrecy here. A strong notion of secrecy demands a vanishing $\left|I(S^n;Z^n,K_1,\cdots,K_r)-nI(S;Z)\right|$, whereas the weak notion of secrecy demands a vanishing $\frac{1}{n}\left|I(S^n;Z^n,K_1,\cdots,K_r)-nI(S;Z)\right|$.

The following theorem provides our result on the secure channel simulation. A slightly stronger version of this theorem can be found in \cite{FullVersion}.
\begin{theorem}
The set of achievable rate tuples includes all non-negative $(R_0,R_{12},R_{21})$, for which there exists $p(f_{[1:r]},x_{[1:2]},y_{[1:2]}, z, s)$ such that equations \eqref{eqT1}-\eqref{eqT4}, the Markov constraints
\begin{align}
X_{[1:2]},Z&,Y_{[1:2]},S\sim p(x_{[1:2]},z)p(y_{[1:2]}|x_{[1:2]})p(s|x_{[1:2]}y_{[1:2]}z),\n
F_i-&F_{[1:i-1]}X_1-X_2Z, \ \mbox{if $i$ is odd,} \n
F_i-&F_{[1:i-1]}X_2-X_1Z, \ \mbox{if $i$ is even,} \n
&Y_1-F_{[1:r]} X_1-X_2Y_2Z,\n
&Y_2-F_{[1:r]} X_2-X_1Y_1Z,\n
&S-X_{[1:2]}Y_{[1:2]}Z-F_{[1:r]}\label{eqM1}
\end{align}\normalsize
and the following additional constraint (for all $i\in [1:r]$) are satisfied.
\begin{align}
I(F_{[1:i]};SZ)+I(X_1;X_2|F_{[1:i]})&< R_0+I(X_1;X_2).\label{eqSE1}
\end{align}\normalsize
\end{theorem}

\emph{Discussion.} The above theorem implies the achievability part of the result of \cite{Tyagi} in the case of two terminals. Consider the special case of $Y_1=Y_2=S=Y=g(X_1,X_2)$, $Z=\emptyset$, $r=2$, $F_1=X_1$, $F_2=X_2$, $R_0=0$, $R_{12}=\infty$ and $R_{21}=\infty$. It shows that a function $Y=g(X_1,X_2)$ can be generated securely at both terminals if $H(Y)<I(X_1;X_2)$. Further if we have a preshared secret key at rate $R_0$, this condition reduces to $H(Y)<I(X_1;X_2)+R_0$.

Next, consider the special case of $Y_2 = \emptyset$ and $Y_1=g(X_1,X_2)$, i.e. only one terminal is interested in computing a function. As before assume $Z=\emptyset$, $r=2$, $R_{12}=\infty$ and $R_{21}=\infty$. In this case we can choose $F_1=\emptyset$ and $F_2=X_2$. This gives us the constraint $I(X_2;Y_1)<I(X_2;X_1)+R_0$. When $R_0=0$ we get a result already known from \cite{Tyagi}.

Another special case is when $H(Y_2|Y_1)=0$ and $Y_1=g(X_1,X_2)$, i.e. the function computed by the second terminal is a function of the one computed by the first terminal. Further assume $S=Y_1$, meaning that we would like to make sure that the eavesdropper learns nothing about $Y_1$. As before we are not charging the public discussion, i.e.  $R_{12}=\infty$ and $R_{21}=\infty$. Assume further that $R_0=0$. It is shown in Corollary 4 of \cite{Tyagi2} that secure computation is possible if and only if $H(X_1, X_2 | Y_1) > H(X_2| X_1) + H(Y_2| X_2) + H(X_1| Y_1, X_2)$. Observe that this condition is equivalent with $I(X_1;X_2)>I(X_2Y_2;Y_1)$. To achieve it we can set $F_1=\emptyset$, $F_2=X_2$, $F_3=Y_2$. 

\begin{proof}
We use the OSRB technique as above and create a single i.i.d. copy of $(F_1^n,\cdots, F_r^n, X_{[1:2]}^n, Y_{[1:2]}^n, Z^n, S^n)$, as well as bin indices $\omega$, $K_i$ and $B_i$ just as above. To impose the secrecy constraint, it suffices to ensure that $(S^n, Z^n)$ is nearly independent of $(B_{1:r}, K_{1:r})$, the public messages and the preshared randomness variables. This implies that for almost all choices of $B_{1:r}=b_{1:r}$, the mutual information $I(S^n; Z^n, K_{1:r}|B_{1:r}=b_{1:r})$ is asymptotically zero. To accomplish this we impose a stronger constraint that implies $B_{1:r}$, $K_{1:r}$ and $(S^n, Z^n)$ are asymptotically mutually independent. Using Thm. 1 of \cite{OSRB} (after removing redundant equations arising because the random variables we are binning are nested) we can write the condition as (see the full version for details \cite{FullVersion}):
\begin{align*}\sum_{t=1}^i (R_t+\tR_t)&<H(F_{[1:i]}|S,Z)&\forall i\in[1:r].
\end{align*}\normalsize
The Reliability and Independence constraints would not change. Applying a Fourier-Motzkin elimination, we get the region given in the statement of the theorem. See \cite{FullVersion} for tricks to do the elimination efficiently.
\end{proof}

\section{Secure Channel Simulation And Secret Key Generation} \label{section:KeyGeneration}
When $R_{12}=R_{21}=\infty$, $Z=\emptyset$ and $R_0=0$, Tyagi et al. have shown that secure computing of a common function $Y_1=Y_2=Y=g(X_1,X_2)$ is possible if and only if $H(Y)<I(X_1;X_2)$. The mutual information $I(X_1;X_2)$ is the secret key capacity of the corresponding source model problem. Thus $H(Y)$ cannot exceed $I(X_1;X_2)$ since $Y$ itself can serve as a secret key. Thus the non-trivial part is the achievability part. The authors in \cite{Tyagi} also show that the terminals can compute $Y$ while simultaneously creating a secret key of rate $I(X_1;X_2)-H(Y)$ that is mutually independent of $Y$ and the public discussion. Therefore the function can be augmented by a residual secret key to yield an optimal SK generation scheme. But what if $Z$ is not a constant rv? The SK capacity is not known in this case. The best known lower bound is given in \cite{SKpaper}. Note that the public discussion was not charged in \cite{SKpaper}. Thus it would be desirable to prove a theorem that unifies these results.

 In our work Alice and Bob generate $Y_1$ and $Y_2$ which are not necessarily equal. Let us first assume that $Y_1=Y_2=Y$. However unlike \cite{Tyagi}, rv $Y$ is not necessarily a function of $(X_1, X_2)$; the conditional pmf $p(y|x_1, x_2)$ can be arbitrary. Setting $S=Y$ guarantees that Eve does not learn about $Y$ more than $I(Y;Z)$. Thus, Alice and Bob can extract a secret key of rate $H(Y|Z)$ (by taking a hash or random bin of their $Y^n$ sequences). In order to augment this key with an additional secret key, Alice and Bob use a code that enables them to simultaneously create a secret key $T$ that is independent of $Y^n, Z^n$ and the public discussion. In this case it is desirable to know if they can create a key of rate ``secret key capacity minus $H(Y|Z)$".

 But how about the general case of $Y_1\neq Y_2$?  Here we cannot use either $Y_1$, $Y_2$ or an $S$ as part of a secret key since neither is available at both parties. The natural extension is to imagine a fourth party, Charles, who is getting $S^n$. Alice and Bob want to generate $Y_1^n$ and $Y_2^n$ while protecting Charles against Eve (by making sure that Eve does not learn anything new about $S^n$). Here Alice and Bob also create a secret key $T$ that is secure against both Eve and Charles, i.e. $I(T;S^n,Z^n,K_1, \cdots, K_r)\rightarrow 0$ as $n$ converges to infinity. In other words, we would like the key to be independent of $S^n, Z^n$ and the public discussion under a strong notion of secrecy. We use $R_{SK}$ to denote the rate of the generated secret key. In the special case of $S=Y_1=Y_2=Y$, this problem reduces to the one discussed in the above paragraph.

 Let us begin with the lower bound of \cite{SKpaper}: for any set of random variables $F_1, F_2, ..., F_r$ such that $F_i-F_{[1:i-1]}X_{(i)_2}-ZX_{(i+1)_2}$ form a Markov chain, and for any $a\in [1:r]$, the secret key rate \begin{align*}&\sum_{i=a}^r \big(I(F_i;X_{(i+1)_2}|F_{[1:i-1]})-I(F_i;Z|F_{[1:i-1]})\big)=\\&I(X_1;X_2|F_{[1:a-1]})-I(X_1;X_2|F_{[1:r]})-I(F_{[1:r]};Z|F_{[1:a-1]})\end{align*}
is achievable. The choice of $a=1$ is the best choice in the lower bound when for any $a'\in[1:r]$
\begin{align}\sum_{i=1}^{a'} \big(I(F_i;X_{(i+1)_2}|F_{[1:i-1]})-I(F_i;Z|F_{[1:i-1]})\big)>0, \label{eqn:LL1}
\end{align}\normalsize
otherwise we can replace $a=1$ with $a'$ to get a strictly larger inner bound. To convey the ideas in the simplest way we restrict ourselves to the lower bound when the choice of $a=1$ is optimal, and state the following theorem. A stronger version of this theorem can be found in \cite{FullVersion}.
\begin{theorem}
Take an arbitrary rate tuple $(R_0, R_{12},R_{21})$ for which there exists $p(f_{[1:r]},x_{[1:2]},y_{[1:2]}, z, s)$ such that equations \eqref{eqT1}-\eqref{eqT4}, the Markov constraints given in \eqref{eqM1} and Eq. \eqref{eqSE1} hold. 
Then a secret key of rate $R_{SK}$ can be simultaneously created during the secure channel simulation protocol if
\begin{align}
R_{SK}&<R_0+\sum_{i=1}^r \big(I(F_i;X_{(i+1)_2}|F_{[1:i-1]})-I(F_i;ZS|F_{[1:i-1]})\big)\nonumber\\&=R_0+I(X_1;X_2)-I(X_1;X_2|F_{[1:r]})-I(F_{[1:r]};ZS).\label{eqn:SK1}
\end{align}\normalsize
\end{theorem}
\begin{remark} When $S=Y_1=Y_2=\emptyset$, $R_0=0$ and $R_{12}=R_{21}=\infty$ we get back the lower bound of \cite{SKpaper} for the case of $a=1$. Eq. \eqref{eqSE1} reduces to \eqref{eqn:LL1} in this case which is automatically satisfied when $a=1$ is an optimal choice. In \cite{FullVersion} we provide a complete generalization.
\end{remark}
\begin{remark} When $S=Y_1=Y_2=Y=g(X_1,X_2)$ and $R_{12}=R_{21}=\infty$ we can set $F_1=X_1$ and $F_2=X_2$ to get achievable secret key rate $[I(X_1;X_2)-I(X_1X_2;Z)]-H(Y|Z)$. When $Z=\emptyset$ we get $I(X_1;X_2)-H(Y)$, indicating that this choice is optimal. However this choice for $F_i, i\in [1:r]$ is not necessarily optimal when $Z$ is not constant.
\end{remark}
\begin{proof} We follow the same scheme as in the previous case, at the end of which we create $T$ as the bin index of a random binning of $F^n_{[1:r]}$ (with the number of bins equal to $2^{nR_{SK}}$). Since $F^n_{[1:r]}$ is available to both parties at the end of the protocol, both parties can agree on $T$ with high probability (see \cite{FullVersion} for rigorous statements). Thus we need conditions that imply $T$ is independent of $(B_{[1:r]}, K_{[1:r]}, S^n, Z^n)$. It suffices to make sure that $T$, $S^n, Z^n$ and $B_{[1:r]}$ and $K_{[1:r]}$ are mutually independent. Using Thm. 1 of \cite{OSRB} (after removing redundant equations) we can write the conditions as (see the full version for details \cite{FullVersion}):
\begin{align*}R_{SK}+\sum_{t=1}^r (R_t+\tR_t)&<H(F_{[1:r]}|S,Z),\\
\sum_{t=1}^{a'} (R_t+\tR_t)&<H(F_{[1:i]}|S,Z),\forall a'\in[1:r].
\end{align*}\normalsize
Applying a Fourier-Motzkin elimination, we get the Eq. \eqref{eqn:SK1} as well as the following additional constraints for any $a'\in[1:r]$
\begin{align*}R_0+\sum_{i=1}^{a'} \big(I(F_i;X_{(i)_2}|F_{[1:i-1]})-I(F_i;ZS|F_{[1:i-1]})\big)>0.
\end{align*}\normalsize
The above constraint is identical to the one given in Eq. \eqref{eqSE1}.
\end{proof}

\section*{Acknowledgment}
The authors would like to thank Prakash Narayan and Himanshu Tyagi for discussions on the secure function computation problem.


\begin{thebibliography}{10}
\bibitem{Ahlswede}
R. Ahlswede and I. Csiszar, \newblock{``Common Randomness in Information Theory and Cryptography. Part I: Secret sharing,"} IEEE Trans. IT, 39 (4), 1121--1132, 1993.

\bibitem{Wyner}
A. D. Wyner, \newblock{``The wire-tap channel,"} Bell Syst. Tech. J., 54 (8), 1355--1387, 1975.

\bibitem{Tyagi}
H. Tyagi, P. Narayan and P. Gupta, \newblock{``When Is a Function Securely Computable?,"} IEEE Trans. IT, 57(10), 6337-- 6350, 2011.

\bibitem{Tyagi2}
H. Tyagi, \newblock{``Distributed Computing With Privacy"}, IEEE Symposium On Information Theory (ISIT) 2012.

\bibitem{me2}
M.~H.~Yassaee, A.~Gohari and M.~R.~Aref,
\newblock{``Channel Simulation via interactive Communications,"} IEEE Symposium On Information Theory (ISIT) 2012, pp. 3058-3062.

\bibitem{OSRB}
M.~H.~Yassaee, M.~R.~Aref and A.~Gohari,
\newblock{``Achievability Proof via Output Statistics of Random Binning,"} IEEE Symposium On Information Theory (ISIT) 2012, pp. 1049-1053.

\bibitem{SKpaper}
A. Gohari and V. Anantharam, \newblock{``Information-theoretic key agreement of multiple terminals: part I,"} IEEE Trans. IT, 56(8): 3973-3996 (2010).

\bibitem{FullVersion}
A.~Gohari, M.~H.~Yassaee and M.~R.~Aref,
\newblock{``Secure Channel Simulation,"} To be available on arXiv.


\end{thebibliography}
\end{document}